\pdfoutput=1
\documentclass[prd,preprint,floats,floatfix,aps,nofootinbib,preprintnumbers]{revtex4}
\usepackage{graphics,graphicx,psfrag}
\usepackage{amsmath,amssymb}
\usepackage[sort&compress]{natbib}
\usepackage{bm}	
\usepackage{verbatim}
\usepackage{dsfont}
\usepackage{multirow}
\usepackage{subfigure}
\usepackage{slashed}
\usepackage{ulem}
\usepackage{color}
\usepackage{ulem}
\usepackage[linktocpage=true,bookmarksnumbered=true,colorlinks=true,plainpages,linkcolor=blue,citecolor=red]{hyperref}
\usepackage[utf8]{inputenc}  

\def\tq{\tilde{q}}
\def\issue(#1,#2,#3){{\bf #1}, #2 (#3)}

\catcode`\@=11
\def\lsim{\mathrel{\mathpalette\@versim<}}
\def\gsim{\mathrel{\mathpalette\@versim>}}
\def\@versim#1#2{\vcenter{\offinterlineskip
\ialign{$\m@th#1\hfil##\hfil$\crcr#2\crcr\sim\crcr } }}
\catcode`\@=12

\catcode`@=12

\newcommand{\newc}{\newcommand}
\newc{\wt}{\widetilde}
\newc{\ra}{\rightarrow}
\usepackage{hyperref}
\def\beq {\begin{equation}}
\def\eeq {\end{equation}}
\def\bi {\begin{itemize}}
\def\ei {\end{itemize}}
\def\bea {\begin{eqnarray}}
\def\eea {\end{eqnarray}}

\def\issue(#1,#2,#3){{\bf #1}, #2 (#3)}

\hypersetup{
   colorlinks=true,       
   linkcolor=blue,        
   citecolor=red,         
   filecolor=magenta,      
   }

\begin{document}


%
\title{ Clockworking FIMPs }
\vspace*{0.25in}   

\author{Andreas Goudelis$^{1,2}$, Kirtimaan A. Mohan$^{3}$, Dipan Sengupta$^{3}$}
\email{andreas.goudelis@lpthe.jussieu.fr,  kamohan@pa.msu.edu, dipan@pa.msu.edu}

\affiliation{\vspace*{0.1in}$^1$Sorbonne Universite, CNRS, Laboratoire de Physique Theorique et Hautes Energies,\\
 LPTHE, F-75252 Paris, France}
\affiliation{\vspace*{0.1in}$^2$Sorbonne Université, Institut Lagrange de Paris (ILP), \\
75014 Paris, France}
\affiliation{\vspace*{0.1in}
 {$^3$Department of Physics and Astronomy\\
Michigan State University \\
 567 Wilson Road, East Lansing U.S.A.\\
}}

\begin{abstract} 
We study freeze-in dark matter production in models that rely on the Clockwork mechanism to suppress the dark matter couplings to the visible sector. We construct viable scalar and 
fermionic dark matter models within this ``Clockwork FIMP'' scenario, with several subtleties that need to be taken into account revealed in the model-building process. We also provide analytic, semi-analytic and numerical results for the diagonalization of Clockwork-type mass matrices and briefly discuss the LHC phenomenology of the corresponding scenarios. 

\end{abstract}
\maketitle

\tableofcontents

\section{Introduction}

The non-observation of TeV-scale new physics at the Large Hadron Collider (LHC) has prompted a critical re-investigation of the naturalness problem of the Standard Model (SM) as an effective theory. A number of interesting proposals have appeared during the last decade, including neutral naturalness \cite{Arkani-Hamed:2016rle,Craig:2014aea}, hidden sector models \cite{Han:2007ae,Abel:2008ai}, twin Higgs \cite{Chacko:2005pe} or the dynamical relaxation of the weak scale in models involving axions, the so-called \textit{relaxion} scenario \cite{Graham:2015cka}. The relaxion picture, in particular, has found an interesting UV motivation in the context of renormalizable quiver theories of scalars which appear as pseudo-Nambu-Goldstone bosons in the low-energy effective action \cite{Kaplan:2015fuy,Choi:2015fiu,Giudice:2016yja,Craig:2017cda,Giudice:2017fmj,Ibanez:2017vfl,Teresi:2018eai,Lee:2017fin}. By employing specific symmetry breaking patterns, such configurations can generate an exponential scale separation between the fundamental theory and the low-energy effective theory. These scenarios, which have been dubbed ``Clockwork'' models, can be further motivated from the deconstruction of extra dimensional models, supersymmetry and string theory\cite{Craig:2017cda,Giudice:2017fmj,Ibanez:2017vfl,Teresi:2018eai}.

However, the Clockwork mechanism can be viewed as a more generic (albeit, not without limitations \cite{Craig:2017cda}) framework to naturally generate exponential hierarchies between different scales or couplings. As such, it has also been exploited in the context of inflation \cite{Kehagias:2016kzt,You:2017kah,Im:2017eju}, neutrino physics \cite{Ibarra:2017tju,Park:2017yrn}, composite Higgs models \cite{Ahmed:2016viu}, flavor \cite{Patel:2017pct} and axion physics \cite{Farina:2016tgd,Coy:2017yex,Agrawal:2017eqm,Bonnefoy:2018ibr}, long story short, in several situations involving large scale separations or small couplings. Besides, the Clockwork idea could also be invoked in order to naturally explain different aspects of dark matter (DM) phenomenology. As an example, in most WIMP scenarios dark matter stability is ensured through the -- by hand -- imposition of some discrete or continuous symmetry (for an interesting overview \textit{cf e.g.} \cite{Hambye:2010zb}). The authors of \cite{Hambye:2016qkf}, instead, proposed a model in which the dark matter particles \textit{can} decay, but only through Clockwork-suppressed couplings and can, hence, be rendered naturally stable over cosmological timescales.

At the same time, the non-observation of dark matter particles in direct \cite{Aprile:2017iyp,Tan:2016zwf,Akerib:2016vxi,Amole:2017dex}, indirect \cite{Ahnen:2016qkx,Abdallah:2016ygi,Giesen:2015ufa} and collider searches \cite{Aad:2015baa,Khachatryan:2016nvf} has motivated the study of dark matter generation mechanisms alternative to the standard thermal freeze-out picture. One such scenario that has received particular attention is the so-called ``freeze-in'' mechanism \cite{McDonald:2001vt,Hall:2009bx} in which the dark matter particles only interact extremely weakly with the visible sector, with the corresponding DM candidates being dubbed ``Feebly Interacting Massive Particles'' (FIMPs).

The most conventional freeze-in scenarios require couplings between dark matter and other bath (in this context, in thermal equilibrium with the SM) particles of the order of $10^{-10} - 10^{-13}$ which may appear unnatural from an IR standpoint. An interesting question is, then, whether the Clockwork idea could be employed in order to explain the smallness of these couplings. A first attempt in this direction was presented in \cite{Kim:2017mtc} and then in \cite{Kim:2018xsp}, in which the authors proposed a Higgs portal scalar dark matter model with Clockwork-suppressed interactions between dark matter pairs and the SM Higgs boson. Within the assumptions adopted in these works, this scenario requires a FIMP mass within the range of a few MeV.

In this paper, we explore alternative possibilities for the ``Clockwork FIMP'' idea. We propose ways to extend the dark matter mass range for the case of scalar dark matter produced through Higgs portal-type interactions and we further construct a model of fermionic Clockwork FIMP. In both cases, we investigate the requirements for successful freeze-in dark matter production, point out various subtleties that appear in the model-building process and discuss the potential LHC phenomenology associated with these frameworks. We do not contemplate on possible UV completions that can generate the Clockwork particle landscape. Rather, we focus our attention on the necessary and salient features of building a Clockwork FIMP model. Lastly, we provide a detailed derivation of the diagonalization of ``Clockwork-type'' matrices and deformations thereof in our Appendices \footnote{A derivation of the eigenvalues and eigenvectors of the original Clockwork can be found in \cite{Farina:2016tgd}.}.

The rest of the paper is organized as follows: in Section \ref{sec:freezein} we recall some key features of the freeze-in dark matter production mechanism that will be of relevance in what follows. Sections~\ref{sec:clockwork} and~\ref{sec:fclockwork} contain the main results of this paper. In each case, after summarizing the Clockwork mechanism we present two concrete models of FIMP dark matter, one scalar (Section \ref{sec:clockwork}) and one fermionic (Section \ref{sec:fclockwork}), we study their viable parameter space and discuss alternative constructions that could be considered. In Section \ref{sec:LHC} we comment on the LHC phenomenology of our main scenarios and in Section~\ref{sec:conclusions} we conclude. Two Appendices follow, in which we describe in detail the procedure to diagonalize the scalar (Appendix \ref{sec:diagonalize}) and fermion (Appendix \ref{sec:fcwdiag}) Clockwork mass matrices.
 
\section{Freeze-in dark matter production}\label{sec:freezein}

In its simplest form, the freeze-in picture relies on two main assumptions for the dark matter particles $\chi$ (for a recent review \textit{cf} \cite{Bernal:2017kxu}). The first one is that early enough during the cosmic evolution, they where absent from the primordial plasma. The second is that they only interact extremely weakly with Standard Model particles or any other particle species that is in thermal equilibrium with the SM thermal bath. The dark matter annihilation rate through a reaction of the type $\chi + X \rightarrow a + b$ scales as $n_\chi n_X \times \left\langle \sigma v \right\rangle$, where $n_i$ is the number density of species $i$ and $\sigma$ the reaction cross-section. Then, the combination of the two freeze-in requirements (small $n_\chi$ and small couplings to the visible sector) implies that any quantity of dark matter particles produced will not annihilate back. In other words, when solving the Boltzmann equation for $\chi$, we can ignore its annihilation term and, as long as $\chi$ is stable, only take into account the integrated collision term corresponding to dark matter production. The latter can occur either from decay or from scattering processes of SM or BSM particles.

In the freeze-in picture, dark matter production typically starts at some temperature $T_R$, called the ``reheating temperature'', and can peak either at much lower temperatures (``IR-dominated'' freeze-in) \cite{Hall:2009bx} or close to $T_R$ (``UV-dominated'' freeze-in, \textit{cf e.g.} \cite{Elahi:2014fsa}). The reheating temperature can find a more precise meaning in the context of inflation, being defined through the inflaton decay rate $\Gamma_\phi$ as $T_R \sim \Gamma_\phi M_{\rm Pl}$, where $M_{\rm Pl}$ is the Planck mass. For the purposes of this work, $T_R$ is simply a parameter representing the temperature at which dark  matter production is assumed to start. Besides, in the scenarios we study in this paper freeze-in is IR-dominated, so the dependence of the predicted dark matter abundance on $T_R$ is very mild.

As already mentioned in the introduction, in typical freeze-in scenarios the observed relic abundance can be reproduced for FIMP couplings to the bath particles of the order of $10^{-10}$ -- $10^{-13}$. One possibility in order to naturally explain such values is to assume that dark matter production is driven from operators involving ${\cal{O}}(1)$ couplings but which are suppressed by some large mass scale, \textit{cf e.g.} \cite{Mambrini:2013iaa}. Another possibility, which is also the direction we follow in this work, is by invoking symmetries and symmetry breaking patterns that necessarily lead to some couplings being highly suppressed. Note that the dark matter particles should not possess substantial interactions with \textit{any} sector that thermalizes with the Standard Model, since this would necessarily lead to the thermalisation of the dark matter particles themselves. This feature, as trivial as it might appear, imposes additional restrictions on FIMP dark matter model-building.

As simple as the freeze-in idea might be, in practice calculations can get fairly cumbersome if a large number of processes contribute to dark matter production. As we will see in the following Sections Clockwork FIMP scenarios do, indeed, tend to involve a large number of BSM fields and processes through which dark matter particles can be produced. In order to account for all of them, we employ the latest version of the {\tt micrOMEGAs} dark matter code \cite{Belanger:2018ccd} which has been recently upgraded in order to compute the abundance of FIMP dark matter candidates in freeze-in scenarios. The implementation of all models in {\tt micrOMEGAs5.0} has been performed through the {\tt FeynRules} package \cite{Alloul:2013bka}.

Let us, finally, also comment on one last point concerning the distribution functions of bath particles in the early Universe. In most existing studies of freeze-in scenarios, it is assumed that the bath particles follow a Maxwell-Boltzmann distribution. In \cite{Belanger:2018ccd} it was shown that dropping this assumption can affect the estimated dark matter abundance by factors up to a few. We have, however, found that dropping the Maxwell-Boltzmann approximation increases the CPU requirements to such an extent that a systematic study of our models is made prohibitive. For this reason, throughout this work we will assume Maxwell-Boltzmann distribution functions for all bath particles. Whereas we expect that our results may change by factors of a few, our qualitative results remain unaltered.

\section{A scalar Clockwork FIMP}\label{sec:clockwork}

We now move on to discuss the Clockwork mechanism and how it can be employed in order to naturally generate the small couplings required for successful freeze-in dark matter production. We start with the scalar case, first recalling the basic ingredients of the mechanism. It goes without saying that a more detailed discussion of the general features of the Clockwork idea can be found in the original references \cite{Kaplan:2015fuy,Choi:2015fiu,Giudice:2016yja}. Here we simply highlight some elements in order to make the discussion as self-contained as possible and to motivate some of our choices. We then study a concrete realisation of a Higgs portal scalar dark matter model and comment on alternative model-building possibilities.

\subsection{The scalar Clockwork mechanism}
\label{scalarcw}
A scalar Clockwork model can be constructed starting from a global $\prod_{i=0}^{N} \otimes U(1)_{i}$ symmetry in some theory space. The $U(1)_{i}$ factors are spontaneously broken at respective scales $f_i$, which for simplcity we take to be equal to a common scale $f$, generating $N+1$ massless Goldstone bosons below this scale. 

The  mechanism, then, is reminiscent of an Ising model with nearest neighbour interactions between lattice sites: the global $U(1)^{N+1}$ is further softly broken by $N$ spurion-like mass parameters $m_{j}^{2}$, each of which is taken to carry a charge
\begin{equation}
Q_{i} = \delta_{ij} - q\delta_{i (j+1)}
\end{equation}
under $U(1)_{i}$. This structure introduces off-diagonal ``interactions'' between scalars charged under adjacent quiver sites. The parameter $q$ is a strength (coupling) characterizing these nearest neighbor interactions. For simplicity, we assume a universal value $m^2$ for all the $m_j$ parameters. In any case, since  $N$ links are explicitly broken, there is one true massless goldstone mode described by the generator
\begin{equation}
\mathcal{Q} = \sum_{j=0}^{N} \frac{\mathcal{Q}_{j}}{q^{j}}.
\end{equation}
By choosing $m^2 << f^2$, we can work within an effective field theory (EFT) in which the only relevant degrees of freedom are the Goldstone bosons, described by the familiar expression
\begin{equation}
U_{j} = e^{i\phi_{j}/f}, \qquad j=0,... , N.
\end{equation}
The EFT Lagrangian reads
\begin{equation}
\mathcal{L}_{SCW} = - \frac{1}{2}\sum_{j=0}^{N} \partial_{\mu} \phi_{j}^{\dagger}\partial^{\mu} \phi_{j} - V(\phi)
\end{equation}
where, expanding up to $\mathcal{O}(\phi^{4})$ in the fields, the scalar potential is given by
\begin{align}
V(\phi) &=  \sum_{j=0}^{N-1} \frac{m^{2}}{2}(\phi_{j} - q\phi_{j+1})^2 + \sum_{j=0}^{N-1} \frac{m^{2}} {24 f^2}(\phi_{j} - q\phi_{j+1})^4 + \mathcal{O}(\phi^{6}) \nonumber \\
           &\equiv \frac{1}{2}\sum_{i,j=0}^{N} \phi_{i}M^{2}_{ij}\phi^{j} +   \frac{m^{2}} {24 f^2}\sum_{i,j=0}^{N} (\phi_{i}M^{2}_{ij}\phi^{j})^2 + \mathcal{O}(\phi^{6}).
           \label{eq:pot}
\end{align}
The $\phi_i$ squared mass matrix $M^2$ can be, as usual, read off the quadratic piece of the potential. It obtains a particular form (\textit{cf} \cite{Giudice:2016yja}) known as the \textit{tridiagonal} matrix and can be diagonalized by a real symmetric orthogonal matrix $O$ as $O^{T}M^{2}O=diag(m^{2}_{a_{0}},\dots,m^{2}_{a_{N}})$, where $a_{0} \dots a_{N}$ are the mass eigenstates. The eigenvalues of $M^2$, for a discrete number of sites, along with the corresponding eigenvectors, can be found by a recursion relation of sequences which we describe in Appendix \ref{sec:diagonalize}. The eigenvectors contain one massless ``zero mode" $a_{0}$ , and a tower of massive states $a_{k}$ (the psueudo-Nambu-Goldstone bosons), dubbed the ``{\it{Clockwork gears}}''. The eigenvalues read
\begin{equation}
m^{2}_{a_{0}}=0,~~m^{2}_{a_{k}}=\lambda_{k}m^{2} ; ~~ \lambda_{k}=q^{2} + 1 -2q\cos\frac{k\pi}{N+1},~~ k=1,\dots,N
\label{scweig}
\end{equation}
whereas the elements of the rotation matrix $O$ are given by
\begin{equation}
O_{j0}=\frac{\mathcal{N}_{0}}{q^{j}},~O_{jk}=\mathcal{N}_{k}\Bigg[q\sin\frac{jk\pi}{N+1} - \sin\frac{(j+1)k\pi}{N+1}\Bigg]; 
\quad j=0,....,N;k=1,,...,N
\label{scweigvec}
\end{equation}
with
\begin{equation}
\mathcal{N}_{0}=\sqrt{\frac{q^{2}-1}{q^{2}-q^{-2N}}} \quad {\rm and} \quad \mathcal{N}_{K} = \sqrt{\frac{2}{(N+1)\lambda_{k}}} 
\end{equation}
 
The masses of the Clockwork gears fill a band of discrete levels, a structure  reminiscent of Kaluza-Klein towers from the deconstruction of compactified extra dimensional set-ups. The mass band starts from $m_{a_{1}} \simeq (q-1)m$ and extends up to $m_{a_{N}} \simeq m_{a_{1}} +\Delta m$, where $\Delta m/m_{a_{1}}=2/(q-1)$. In the large $N$ limit, the mass gap between different levels is
\begin{equation}
\frac{\delta m_{k}}{m_{a_{k}}} \sim \frac{q\pi}{N\lambda_{k}}\sin\frac{k\pi}{N+1}, \quad k=1,\dots,N-1
\end{equation}
 
The crucial feature of the Clockwork setup is the isolation of the goldstone mode for sufficiently large values of $N$ and $q$. If a theory is coupled to the $N$-th site of the Clockwork, the induced interactions with the zero mode are suppressed by $q^{N}$, thus leading naturally to an extremely feeble coupling. This is an essential ingredient for the construction of a FIMP model. We now proceed to the concrete construction of such a model.

\subsection{A model of scalar FIMP}

Although there are numerous ways to implement the freeze-in scenario with a real scalar DM candidate, we will focus on a simple Higgs portal interaction. Such a scenario was also discussed in \cite{Kim:2017mtc,Kim:2018xsp}. Here we expand on this idea and highlight some subtleties that need to be taken into account. 

Our starting point is a standard scalar Clockwork chain, like the one discussed previously, the $N$-th site of which is coupled to the Standard Model via a Higgs portal term as $\kappa |H^{\dagger}H|\phi_{N}^{2}$, where $\kappa$ is a dimensionless coupling that we set to its maximally natural value of $1$~\footnote{In~\cite{Kim:2017mtc}, this coupling was achieved by introducing an additional spurion mass term. We do not concern ourselves with the origin of such a term and only note that it can exist and it is renormalizable.}. Above the electroweak symmetry breaking scale, the Clockwork sector can be diagonalized as described in Appendix \ref{sec:diagonalize}. In this phase, the zero mode $a_0$ (our FIMP candidate) is strictly massless and its interactions with pairs of the Higgs doublet components are suppressed by $\kappa/q^{2N}$.

Once electroweak symmetry is broken, the $N$-th site $\phi_N$ acquires an additional mass contribution through the Higgs vacuum expectation value (vev) as $\kappa v^2 \phi_N^2/2$. The effect of this term on the Clockwork zero-mode eigenvalue is suppressed. This has been checked numerically but can also be gleaned by noticing that the solution to the transcendental relation  Eq.\eqref{rel2} for the massless mode is modified by $q\simeq\exp(i\theta) + \epsilon$, where $\epsilon\sim v^2/(m^2 q^{2N-2}(q-1))$ when $v\ll m$. The FIMP mass therefore scales as $\sim v/\sqrt{2q^{2N-2}(q-1)}$. For example, choosing $q=2$ and $N=5$ we find $m_{a_0} \sim 10$ GeV whereas for $q \simeq 10$ and $N=10$ the $a_0$ mass lies in the sub-keV range. With a bit of hindsight (\textit{cf} also \cite{Kim:2017mtc}), given that the values of $q$ and $N$ for which successful freeze-in can be achieved correspond rather to the latter choice, it would be useful to find a way to raise the FIMP mass. In \cite{Kim:2017mtc}, this was done by adding an additional mass term to the $N$-th site of the Clockwork, which allowed the authors to raise the DM mass to the MeV range. Here we follow an alternative method, which we find to provide even more freedom.

Let us introduce an additional mass term for \textit{all} sites by supplementing the potential in Eq.\eqref{eq:pot} with \footnote{One can consider the additional mass term as background values of spurions emerging
from a term $t^2 f^{2}(U^{q_{j}\dagger} + h.c)$, where $U^{q_j}=exp(i\phi_{j}/f)$. The spurion transformation is then, $Q_{j}[t_{j}^{2}] = q_{j}, j=0,1,2,\dots,N$, with $q_{j}=1~ \forall~ j$ and $t_{j}=t ~\forall~ j$. },
\begin{align}
\sum_{i=0}^{N}\frac{t^2}{2}\phi_i^2\ ,
\end{align}
so that the new mass matrix $M_{t}$ becomes
\begin{equation}
 M_{t} = m\cdot \begin{bmatrix}
 1+t^2/m^2 & -q & 0  &\dots & 0  & 0\\
 -q & 1+ q^{2} +t^2/m^2 & -q & \dots & 0 & 0 \\
 \vdots & \vdots & \vdots & \vdots & \vdots & \vdots  \\ 
 0 & 0 &  0 &\dots & 1+ q^{2} +t^2/m^2 & -q \\
 0 & 0 & 0 & \dots & -q & q^{2} +t^2/m^2
 \end{bmatrix}_{(N+1)\times(N+1)}\ .
\label{scwm}
\end{equation}
This term adds to all the diagonal elements of the mass matrix and has no effect on the form of the transcendental relation Eq.\eqref{rel2}. Hence, the diagonalizing matrix $O$ has the same form as in the original Clockwork setup. On the other hand the eigenvalues, and therefore the squared masses, contain an additional factor of $t^2$. Introduction of the additional parameter $t$, thus, allows us to control the value of the FIMP mass. Note that both the Higgs vev and the additional mass term contribute to the mass of the zero mode and explicitly break the remnant U(1) symmetry of the Clockwork chain.

At this stage, we moreover introduce a deformation of the quartic piece of the Clockwork sector potential: we replace the matrix $M^2$ in the second term of the last line of Eq.\eqref{eq:pot} by a matrix $\tilde{M}^{2}$ the form of which will be given shortly. As we will explain later on, this -- seemingly arbitrary -- choice is motivated by practical considerations.
\\
\\
All in all, our Lagrangian reads
\begin{eqnarray}
\mathcal{L}_{sFIMP}& =& - \frac{1}{2}\sum_{j=0}^{N} \partial_{\mu} \phi_{j}^{\dagger}\partial^{\mu} \phi_{j}- \frac{1}{2}\sum_{i,j=0}^{N} \phi_{i}M^{2}_{ij}\phi^{j} -   \frac{m^{2}} {24 f^2}\sum_{i,j=0}^{N} (\phi_{i}\tilde{M}^{2}_{ij}\phi^{j})^2  \nonumber\\&\,& - \kappa |H^{\dagger}H|\phi_{N}^{2} + \sum_{i=0}^{N}\frac{t^{2}}{2}\phi_{i}^{2}\ .
\label{sfimp}
\end{eqnarray}

After electroweak symmetry is broken, the original Clockwork sector mass matrix is modified to be
\begin{align}
\tilde{M}_{ij}\equiv M_{ij}+ \kappa v^2\delta_{i N}\delta_{jN} .
\end{align}
The exact value of $f$, the breaking scale, is not important here since our choice of quartic couplings ensures that they do not play a part in dark matter production~\footnote{The value of $f$ becomes important if there are off-diagonal interactions between the zero mode and higher gears.}.  The gear masses, driven by the $m$ parameter are chosen to be in the multi-TeV range. This choice also helps evade constraints from the LHC and further ensures that $v \ll m$ and will, thus, have only a small effect on the form of the diagonalizing matrix $O$. Note, however, that we do not neglect this effect in our calculations and numerically determine the exact matrix $\tilde{O}$ that diagonalizes $\tilde{M}^2$.

Interactions between dark matter pairs and the standard model arise through the term $ |H^{\dagger}H|\phi_{N}^{2}$. Once we expand the Clockwork states in terms of their mass eigestates as $\phi_{k} = \sum_{l}O_{kl}a_{l}$, we obtain trilinear and quartic interactions as
\begin{align}\label{eq:sinteractionterms}
\mathcal{L}_{int} & =   \kappa |H^{\dagger}H|\phi_{N}^{2} \nonumber \\
& = \kappa\sum_{l=0,m=0}^{N} O_{Nl}O_{Nm}a_{l}a_{m}(v^2 + 2vh + h^2)/2
\end{align} 
As we already noted, the first term also provides a small contribution to the FIMP mass that is suppressed by a factor $\sim q^{-N}$. The zero mode-zero mode-Higgs interaction is, thus, suppressed by $\sim\frac{\kappa}{q^{2N}}$, while the zero mode-$j$-th gear-Higgs interaction is supressed by $\sim\frac{\kappa}{q^{n}}O_{jN}$. Finally there are gear-gear-Higgs(-Higgs) interactions that are unsupressed by the Clockwork mechanism: at early enough cosmic times all gears are in kinetic and chemical equilibrium with the SM thermal bath.

\begin{figure}
\includegraphics[scale=0.6]{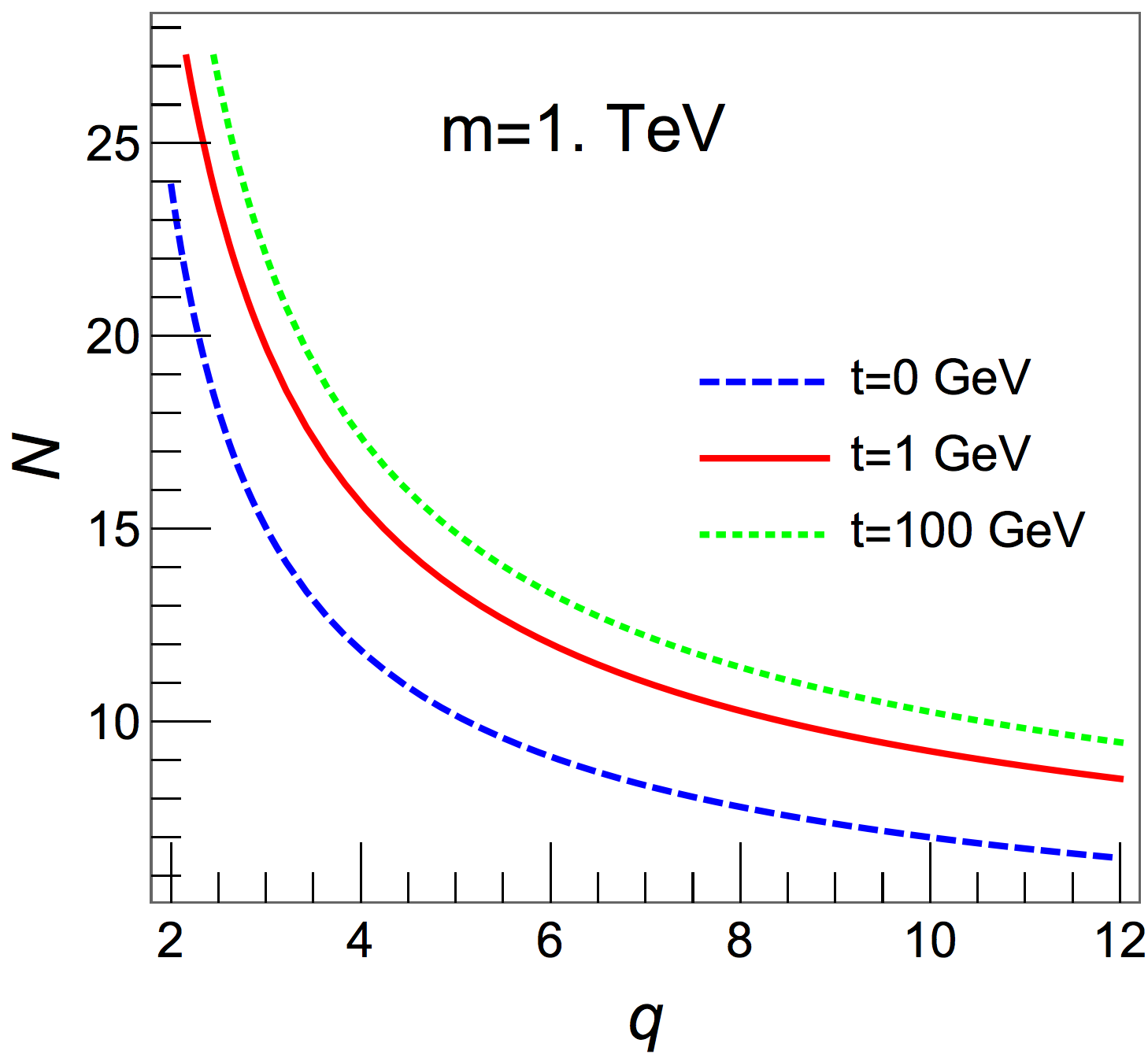}
\includegraphics[scale=0.6]{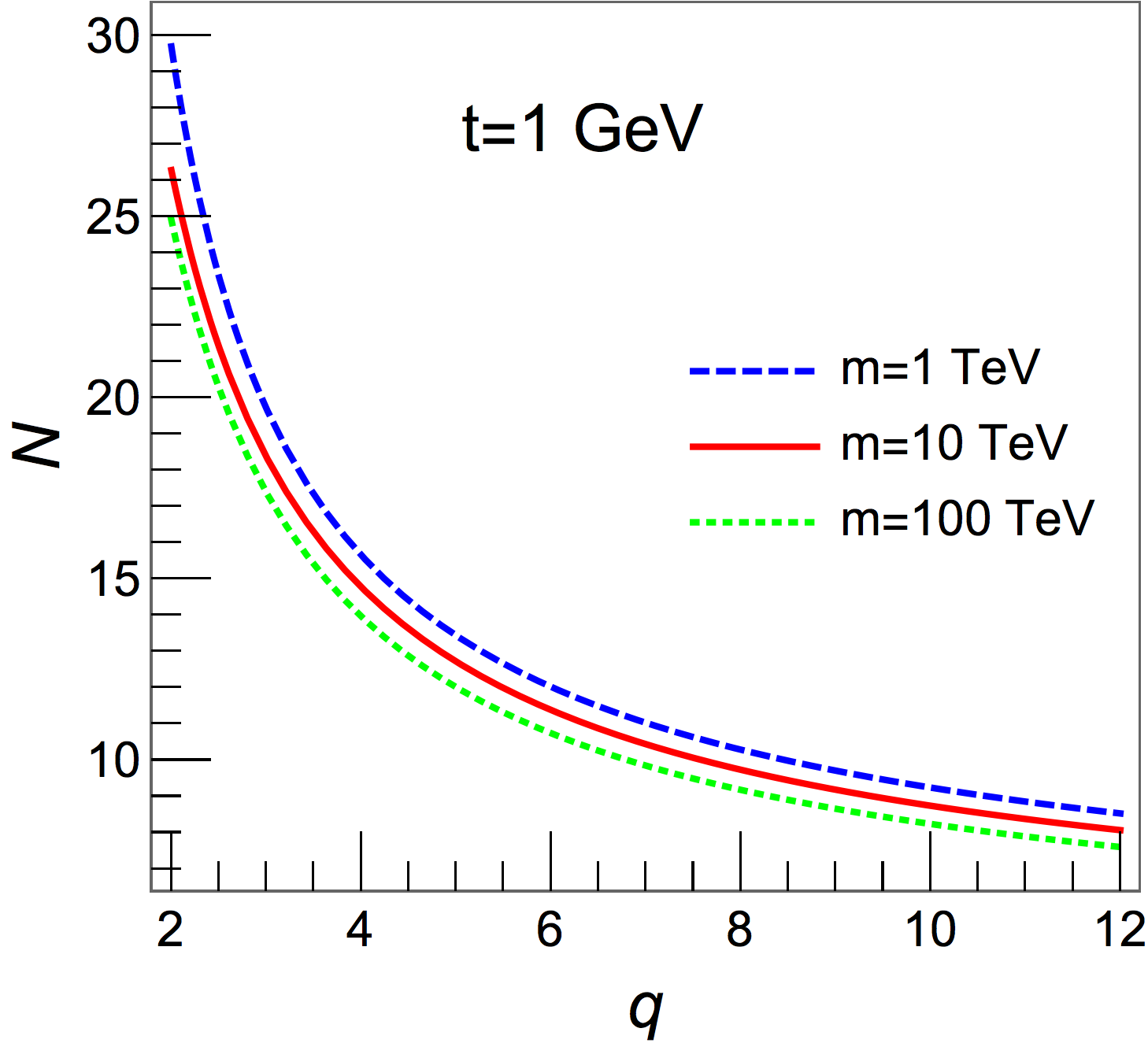}
\caption{Contours indicating the values of $q$ and $N$ that satisfy the observed relic abundance for $\kappa=1$. \textbf{Left:} We choose $m=1$~TeV and show contours for three different choices of the mass parameter $t=\left\{0,1,100\right\}$~GeV. \textbf{Right:} We choose $t=1$~GeV and show contours for three different choices of the mass parameter $m=\left\{1,10,100\right\}$~TeV.
	\label{fig:sfimp}
	}
\end{figure}

In Fig.\ref{fig:sfimp} we show contours in the $q-N$ plane for which the dark matter abundance measured by Planck \cite{Ade:2015xua} can be reproduced in our model according to the freeze-in mechanism. As expected, we see that for small values of $q \sim 2$, this can be achieved for large $N$ values ($\sim 25-30$), whereas $N\sim 10$ is sufficient for larger values of the Clockwork symmetry breaking charge $q\sim 10$. In the left hand-side figure we fix $m=1$ TeV and show contours for three different choices of the $t$ mass parameter, $t=0$, $1$ and $100$ GeV. For $t=0$ the mass of the FIMP is generated entirely from the terms in Eq.\eqref{eq:sinteractionterms}. It ranges from a few GeV at low $(q, N)$ -- a regime in which, however, the DM candidate is highly overabundant\footnote{In fact, in these cases $a_0$ is not even really a FIMP!} -- up to a few keV at larger values. For $t=\{1,100\}$ GeV, on the other hand, the mass of the FIMP receives two contributions and, for large enough $(q, N)$, remains close to the value of $t$. Since the mass of the FIMP increases with increasing $t$, we see that larger values of $q$ and $N$ (\textit{i.e.} further suppression from the Clockwork mechanism) are needed for larger values of $t$ in order to get the correct relic abundance. In the right hand-side figure, we fix $t=1$ GeV and vary the value of $m$. We see that as $m$ decreases, the modification of the Clockwork mass matrix by the Higgs vev grows in prominence requiring larger values of $q$ and $N$ in order to suppress the coupling of the FIMP with the SM. Note that in all cases the dominant DM production is due to the decays of the Clockwork gears into $a_0$ and a Higgs boson.

Let us also comment on our choice of modifying the quartic term from its original form in Eq.\eqref{eq:pot}. Arguably, this choice somehow contradicts the original, symmetry-motivated spirit of the Clockwork mechanism and breaks the connection between the two lines of Eq.\eqref{eq:pot}. The reason why we have opted for this Lagrangian is a practical one: if we were to start with the Lagrangian of Eq.\eqref{eq:pot}, upon EWSB and after diagonalising the scalar sector mass matrix we would pick up quartic interactions between the zero mode and all gears. Although these interactions are Clockwork-suppressed as well, they give rise to a large number of processes contributing to the dark matter relic abundance, making the problem almost intractable from a computational standpoint. Our approximation ensures that despite the modification of the Clockwork mass matrix and eigenbasis by the Higgs vacuum expectation value, there are no quartic interactions between pairs of zero modes and pairs of gears. In fact, in the mass eigenbasis, the quartic term is simply $\sum a_{i}^4$. We can thus completely ignore the quartic terms when calculating the dark matter relic abundance. We emphasize that -- in full generality -- this choice of quartic coupling is not necessary for the freeze-in mechanism to produce the correct relic abundance. An alternative possibility in order to reduce the number of processes contributing to DM production could have been to stick to the original Clockwork scalar potential, Eq.\eqref{eq:pot}, and to send the masses of all gears above the reheating temperature $T_R$. In this case, the abundance of all gears would be exponentially suppressed and the only processes contributing to DM production would be annihilation processes of SM particles through the Higgs portal (or, for appropriate parameter choices, the decay of the Higgs boson into DM pairs). Note that this would be a ``technically natural'' choice, since in the limit $m/f \rightarrow 0$ the symmetry of the theory is increased. Further, we  also ensure that $f \gtrsim m(1+q)$. This condition ensures that the quartic couplings remain perturbative and are not very large\footnote{Recall that the for large $N$ the largest mass is $\sim m(1+q)$.}.

Before moving to the fermion Clockwork, we finally point out that an alternative scalar FIMP scenario could be obtained by introducing an additional singlet scalar $s$ coupled on one hand to the last site of the Clockwork sector and on the other hand to the Standard Model via a Higgs portal interaction. The Lagrangian in this case reads
\begin{equation}
\mathcal{L} = \mathcal{L}_{SCW} + (\partial_{\mu}s)^{2} + \mu^{2}s^2 + \lambda s^{4} +   \kappa s^{2}|H^{\dagger}H| + \xi s^{2}\phi_{N}^{2}  + \sum_{i=0}^{N}\frac{t^{2}}{2}\phi_{i}^{2} 
\label{altscal}
\end{equation}
Here we assume that $s$ does not acquire a vacuum expectation value. Arguably, this model has a simpler structure, as the Clockwork sector is not modified and all the analytic results presented in Appendix \ref{sec:diagonalize} can be used promptly. Once again, the FIMP mass can be controlled via the $t^2$-terms and the relic density can be populated by the decay of heavier gears to the singlet scalar and the zero mode via a suppressed coupling proportional to $1/q^N$. In such a scenario there is the possibility that $s$ is also a dark matter candidate. The relic abundance from $s$ can proceed through usual thermal freeze-out. In order to ensure that the dominant contribution to the relic abundance comes from freeze-in of the FIMP and not from freeze-out of $s$ one must choose $\kappa$ to be of order 1 or larger.

\section{A fermion Clockwork FIMP}\label{sec:fclockwork}

We now turn to the case of the fermion Clockwork mechanism, following the same line of presentation as in the scalar case. We first recollect the general features of fermionic Clockwork constructions and then propose a concrete realization of a fermion Clockwork FIMP scenario.
\subsection{The fermion Clockwork mechanism}
\label{fcwint}
While the scalar Clockwork relies on goldstone symmetry, the fermionic Clockwork is based on chiral symmetry. We introduce $N+1$ right-handed chiral fermions $\psi_{R,j}$, $j=0,\dots,N$ and $N$ left-handed chiral fermions $\psi_{L,i}$, $i=0,\dots,N-1$. The chiral symmetry is broken by $N$ mass parameters $m_{i}$, as well as $N$ linking mass parameters $m q_{i}$ that induce nearest-neighbour interactions. Much like in the scalar case, the linking parameters can be treated as background values of spurions. Then, since  $N$ out of the $N+1$ sites are broken by the mass parameters, a remnant  right-handed chiral fermion remains massless. In general, one can explicitly break the symmetry by adding a Majorana mass term either to the last site or to every site for both the left- and right-handed chiral fermions. Working under this assumption, we can write a Clockwork fermionic lagrangian as \cite{Ibarra:2017tju}, 
\begin{align}
\mathcal{L}_{FCW} & =  \mathcal{L}_{kin} - m\sum_{i=0}^{N-1}(\bar{\psi}_{L,i}\psi_{R,i} - q\bar{\psi}_{L,i}\psi_{R,i+1} + h.c) - \frac{M_{L}}{2}\sum_{i=0}^{N-1}(\bar{\psi}_{L,i}^{c}\psi_{L,i}) - \frac{M_{R}}{2}\sum_{i=0}^{N}(\bar{\psi}_{R,i}^{c}\psi_{R,i}) \nonumber \\
 & =  \mathcal{L}_{kin} -\frac{1}{2}(\bar{\Psi}^{c}\mathcal{M}\Psi + h.c)
\label{fcw}
\end{align}
Instead of diagonalizing the mass matrix in the $\psi_{L}$ and $\psi_{R}$ basis, which would require a biunitary transformation, we diagonalize the mass matrix $\mathcal{M}_{(2N+1)\times(2N+1)}$ in the basis $\Psi_{2N+1} = (\psi_{L,0},\dots,\psi_{L,N-1},\psi^{c}_{R,0},\dots,\psi^{c}_{R,N})$. Here $\psi^{c}$ denotes the charge-conjugated field. Note that this was also the way the fermionic mass matrix was diagonalized in \cite{Ibarra:2017tju}. For simplicity we assume that $M_{L}=M_{R}=m\tilde{q}$. The matrix $\mathcal{M}$ then reads
\begin{equation}
\mathcal{M} = m\cdot\begin{bmatrix}
\tq & 0 & \dots & 0 & 1 & -q & 0  & \dots  & 0\\
0 & \tq & \dots &  0 & 0 & 1  & -q & \dots  & 0 \\
\vdots & \vdots & \vdots & \vdots & \vdots & \vdots &\vdots &\vdots & \vdots \\
0 & 0 & \dots & \tq & 0 & 0 & 0 &\dots & -q  \\
1 & 0  & \dots & 0 &\tq & 0 & 0 & \dots & 0 \\
-q& 1 & \dots & 0 & 0 & \tq & 0 & \dots & 0\\
 \vdots & \vdots & \vdots & \vdots & \vdots & \vdots &\vdots &\vdots & \vdots \\
0 & 0 &\dots &-q & 0 & 0 &0 &\dots&\tq 
\end{bmatrix}_{(2N+1)\times(2N+1)}\ .
\label{fcwm}
\end{equation}
The procedure to diagonalize this matrix is described in Appendix \ref{sec:fcwdiag}. The eigenvalues are found to be
\begin{align}
m_{0} & = m\tq\ , \nonumber\\
m_{k} &= m(\tq - \sqrt{\lambda_{k}}),~k=1,\dots , N\ , \nonumber\\
m_{n+k} &= m(\tq + \sqrt{\lambda_{k}} ),~k=1,\dots , N\ , 
\label{fcweig}
\end{align}
where $\lambda_{k}$ is defined in Eq.\eqref{scweig}. The mass eigenstates, denoted by $\chi_{i}$, are related to the original states $\psi$ through a unitary rotation matrix $\mathcal{U}$
\begin{equation}
\mathcal{U} =
 \begin{bmatrix}
\vec{0} &  \frac{1}{\sqrt{2}} U_{L} &  -  \frac{1}{\sqrt{2}} U_{L} \\
\vec{u}_{R} & \frac{1}{\sqrt{2}} U_{R} &   \frac{1}{\sqrt{2}} U_{R} \\
\end{bmatrix}_{(2N+1)\times(2N+1)}\ ,
\label{fcwrot}
\end{equation}    
such that the eigenstate expansion is $\psi_{i} = \sum_{j}U_{ij}\chi_{j}$. The elements of the unitary transformation matrix $\mathcal{U}$ are given by
 \begin{align}
 \vec{0}_{i} & =  0,~i=1,\dots ,N \nonumber \\
 (\vec{u}_{R})_{i} & = \frac{1}{q^{i}}\sqrt{\frac{q^{2}-1}{q^{2}-q^{-2N}}},~~i=0,\dots , N \nonumber\\
 (U_{L})_{ij}& = \sqrt{\frac{2}{N+1}}\sin\frac{ij\pi}{N+1},~~i,j=1,\dots ,N \nonumber \\
 (U_{R})_{ij} & = \sqrt{\frac{2}{(N+1)\lambda_{j}}}\Bigg[q\sin\frac{ij\pi}{N+1} - \sin\frac{(i+1)j\pi}{N+1}\Bigg], \ i=0,\dots , N,~~j=1,\dots ,N
 \label{fcweigvec}
 \end{align}
Thus $\vec{0}$ and $\vec{u}_{R}$ are column vectors of size $N$ and $N+1$ respectively, while $U_{L}$ and $U_{R}$ are matrices of dimension $N\times N$, and $(N+1)\times N$. In the limit where all the Majorana masses vanish we recover the expressions for the eigenvalues and eigenvectors obtained in Ref.\cite{Giudice:2016yja}, with $U_{L}$, $U_{R}$ being the bi-unitary transformation matrices that diagonalize the left-handed and right-handed chiral fermions. Crucially, the above eigenvectors imply that the Clockwork mechanism is not altered by adding Majorana mass terms to the fermionic Clockwork matrix (the $q^{j}$ suprression of the zero mode is still present), as it only adds a constant diagonal matrix, analogously to the scalar case. An additional interesting feature is that the zero mode may not be lightest mode, depending on the Majorana parameter $m\tq$. However, as long as $\left|m(\tilde{q} - \lambda_{k}) \right| > m\tilde{q}$, the lightest mode is indeed the zero mode. This feature was explored in Ref.\cite{Ibarra:2017tju}.

\subsection{A model of fermionic FIMP}
\label{sec:ffimp}

A fermion FIMP model can be constructed \textit{e.g.} by coupling the Clockwork sector to the Standard Model Higgs boson via a Yukawa interaction. Although this can be also achieved by directly coupling the $N$-th site to the SM left-handed fermions, in order to easily evade flavour constraints we supplement the Clockwork Lagrangian described in Section \ref{fcwint} with additional vector-like leptons $L'=(l_{1},l_{2})$ and $R'=(r_{1},r_{2})$ transforming as $(\mathbf{1},\mathbf{2},\mathbf{-1/2})$ under $SU(3)_c \times SU(2)_L \times U(1)_Y$~\footnote{Introducing leptons and scalars of higher $SU(2)$ representations would also be a conceivable scenario, however, we restrict ourselves to the simplest case.}. We moreover impose a discrete ${\cal{Z}}_2$ symmetry under which the SM is even and all other particles are odd and couple the last site of the Clockwork chain ($\psi_{R,N}$) to the Standard Model Higgs through a Yukawa coupling involving the left-handed exotic lepton doublet. Our Lagrangian is, thus 
\begin{equation}\label{eq:ffimplag}
\mathcal{L}_{fFIMP} = \mathcal{L}_{FCW} + i\bar{L'}\slashed{D}L' +  i\bar{R'}\slashed{D}R' +  M_{D}(\bar{L'}R') + Y\bar{L'}\tilde{H}\psi_{R,N} +~\rm h.c 
\end{equation} 
where $M_{D}$ is the vector-like fermion Dirac mass and $Y$ is the Yukawa coupling.

The vector-like leptons receive their mass from the Dirac mass term in Eq.\eqref{eq:ffimplag}, whereas upon electroweak symmetry breaking the heavy ``neutrinos'' receive an additional contribution from the Yukawa term. This contribution is off-diagonal and induces a mass mixing between the Clockwork gears and the heavy neutrinos. The overall mass matrix is, hence, expanded with respect to its original form, Eq.\eqref{fcwm}, and the diagonalisation must be performed in the extended basis including all BSM neutral fermions in the model. It reads
\begin{eqnarray}
m_{\nu} =
\bordermatrix{
        & l_{1} & r_{1} & \chi_{0} 
        & \chi_{1} & \chi_{2} & \cdots 
        & \chi_{2N} \cr
        l_{1} & 0 & M_{D} & v Y_{0}  
        & v Y_{1} & v Y_{2} &   \cdots
        & v Y_{2N} \cr
        r_{1} & M_{D} & 0 & 0 & 0 & 0 &\cdots & 0 
        \cr
        \chi_{0} & v Y_{0} & 0 & M_0
        & 0 &    0        &\cdots
        & 0 \cr
        \chi_{1} & v Y_{1} & 0 & 0 & M_1   &    0        &\cdots
        & 0 \cr
        \chi_{2} & v Y_{2} & 0 & 0 & 0 & M_2  
        &\cdots
        & 0 \cr
        \vdots & \vdots &\vdots   
        &\vdots & \vdots & \vdots   & \vdots       & \vdots \cr
        \chi_{2N} & v Y_{2N} &  0 & 0 & 0 & 0  &\cdots & M_{2N} 
        \cr
}\;,
\label{eq:general_case}
\end{eqnarray}
where we have used the basis of states where the pure Clockwork sector is diagonal, defined through $\psi_{R,k} = \sum_{l=0}^{N} \mathcal{U}_{kl}\chi_{l}$. Moreover
\begin{align}
&Y_0 = Y(u_R)_N=\frac{Y}{q^{N}}\sqrt{\frac{q^{2}-1}{q^{2}-q^{-2N}}}& \\
&Y_j= Y_{(N+j)}= \frac{Y}{\sqrt{2}}(U_{R})_{Nj}  =Y \sqrt{\frac{1}{(N+1)\lambda_{j}}}\Bigg[q\sin\frac{Nj\pi}{N+1} \Bigg],~~j=1,\dots ,N&
\end{align}
The mass matrix \eqref{eq:general_case} is difficult to diagonalize analytically. In any case, the Lagrangian \eqref{eq:ffimplag} amounts to interactions of the FIMP candidate $\chi_0$ with the SM Higgs boson as well as, due to the mixing between the Clockwork and the vector-like fermions, the SM gauge bosons. All these interactions are suppressed by the Clockwork mechanism, whereas the mixing is governed by the ratio $v/m$. In the limit $v/m \to 0$, all off-diagonal entries (except for $M_D$) vanish, as do all couplings of the Clockwork sector to gauge bosons. In this limit the only interaction of the Clockwork sector to SM particles occurs through the Yukawa coupling. Then, FIMP production can proceed through decays of heavier Clockwork gears into $h \chi_0$ final states, or via SM scattering processes mediated by the Higgs. Interactions between fermion gears $\chi_k$ and the SM are not suppressed by the Clockwork mechanism, although in the limit $v/m \to 0$ interactions with the SM gauge bosons vanish. Hence, much like in the scalar Clockwork case, at early enough cosmic times the gears are in thermal equilibrium with the SM bath thanks to their interactions with the Higgs boson.

Although in principle it is possible to evaluate the relic density taking into account the full mixing of the Clockwork fermions with the heavy neutrino, this is a numerically challenging task~\footnote{Roughly, there are about $2N+1$ processes one needs to consider when we use the assumption $v/m \to 0$, which increases to $3(2N+3)^2$ processes when we consider all mixings.}. In light of the previous remarks, and in order to render the problem computationally tractable, we instead choose to place ourselves in the limit $v/m \to 0$ by fixing $m=100$~TeV. This choice allows us to ignore the interactions of the FIMP with the SM gauge bosons. We have cross-checked that the results obtained under this approximation agree with the full computation at the percent level. Smaller values of $m$ would increase mixing between the Clockwork sector and the vector-like fermion and thus require larger values of $q$ and $N$ to obtain the correct relic.

\begin{figure}
\includegraphics[scale=0.6]{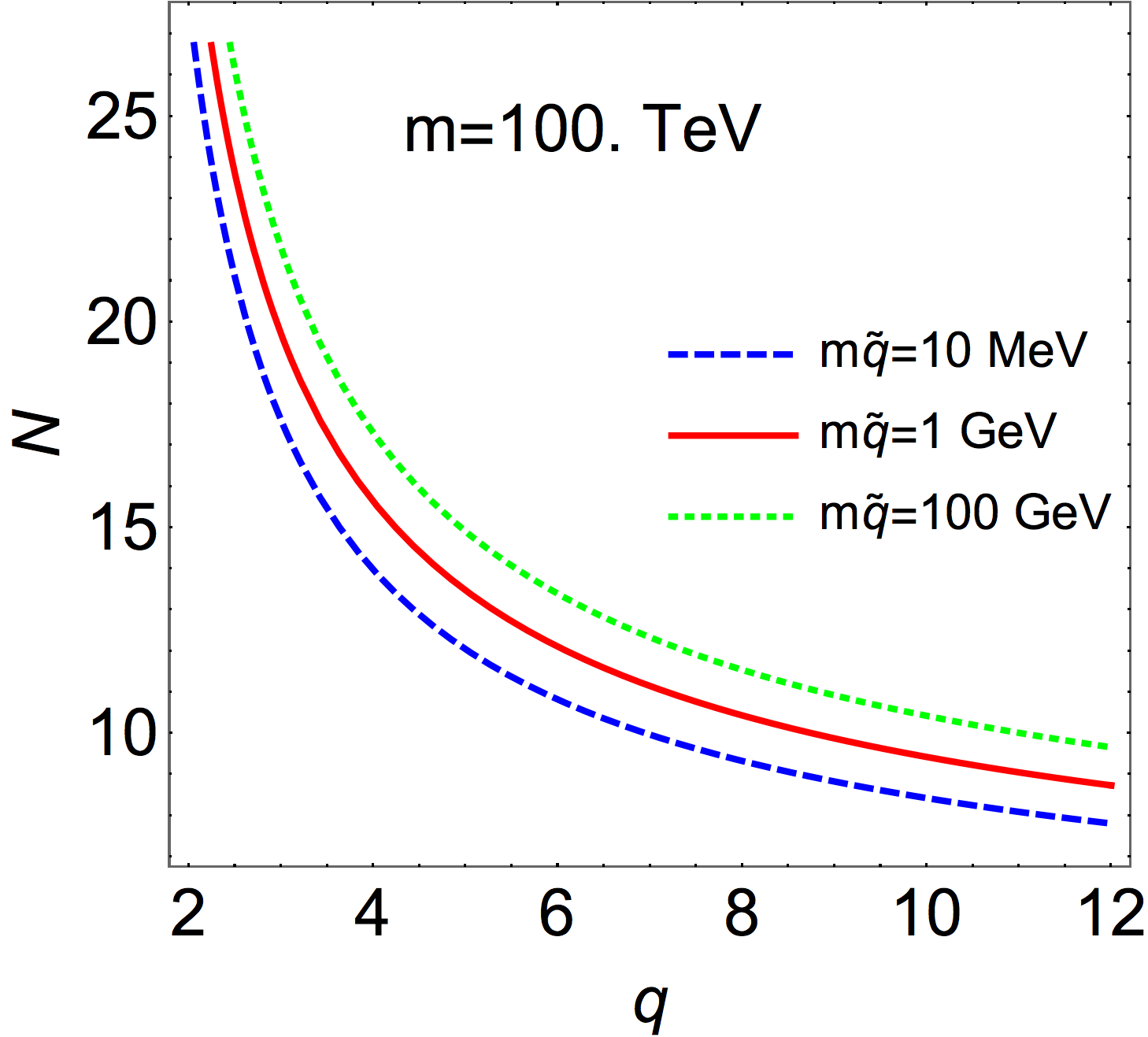}
\caption{Contours denoting the values of $q$ and $N$ that produce the observed relic abundance for the fermionic Clockwork model. Here $m=100$~TeV and the three contours are shown for three choices of $m\tq=\{0.01,1,100\}$ GeV.
	\label{fig:ffimp}
	}
\end{figure}
In Fig.\ref{fig:ffimp} we present contours in the $q-N$ plane for which the observed dark matter abundance can be reproduced, fixing $m=100$ TeV and for three different choices of $m\tq=\{0.01,1,100\}$ GeV. As in the scalar case, we observe that for small values of $q\sim2-3$ large values of $N\sim 20 -25$ are needed, whereas for large values of $q\sim10$ smaller values of $N\sim 8-10$ are sufficient. We can also see that as we increase the value of $m\tq$ the required values of $q$ and $N$ also increase, as smaller coupling values are needed in order to produce the amount of dark matter necessary to explain the Planck observations. In this case, as well, the dominant dark matter production mechanism is the decay of the vector-like leptons (which, for our parameter choices, are the next-to-lightest ${\cal{Z}}_2$-odd particles) into $h \chi_0$ final states.

Let us comment on one subtlety: in typical freeze-in scenarios, and in particular in cases where FIMP production occurs through the decay of a heavier particle, most of dark matter is produced at a temperature which corresponds roughly to one third of the mass of the decaying particle. Then, for a choice $m=100$ TeV, dark matter production peaks at a temperature around 30 TeV, \textit{i.e.} far above the temperature of electroweak symmetry breaking. Strictly speaking, this feature would necessitate computing the total dark matter yield both in the unbroken and in the broken phase of the Standard Model for different temperature windows. However, since we work in the limit $v/m \to 0$ and we neglect the Yukawa-induced mass mixing between the gears and the vector-like leptons, the masses of all BSM particles (including the zero mode and the vector-like fermions) are fully controlled by parameters with no relation to EWSB and the only differences between the broken and the unbroken phase of the Standard Model which could be of some relevance for dark matter production are the mass of the Higgs boson and whether it is complex or real. At the end of the day, both of these factors are of minor importance and, for simplicity, we perform our calculations assuming electroweak symmetry is broken.

Note, also, that an alternative option to couple a fermionic Clockwork chain to the Standard Model could be to introduce an additional singlet  $s$ that couples  to the last site of the right-handed sector of the quiver\footnote{One could also consider introducing an additional left-handed fermion.}. In this case, the introduction of vector-like fermions would not be needed and the Lagrangian would read
\begin{equation}
\mathcal{L} = \mathcal{L}_{FCW} + (\partial_{\mu}s)^{2} + \mu^{2}s^2 + \lambda s^{4} + \kappa s^{2}|H^{\dagger}H| +   Y_{s}s(\bar{\psi}_{R,N}^{c} \psi_{R,N})
\label{altfer}
\end{equation}
Interactions with the SM are mediated by the Higgs portal term $s^2|H^\dagger H|$. As mentioned earlier, as long as $s$ does not acquire a vacuum expectation value, there is no modification to the Clockwork mass matrices. The coupling of $s$ to the FIMP candidate ($\chi_0$) and the heavier gears is therefore suppressed exactly by $1/q^N$ . In this case the FIMP obtains a mass via the Majorana mass terms. 

\section{Comments on collider phenomenology}\label{sec:LHC}

While the primary objective of this study is to assess the viability of freeze-in dark matter in Clockwork constructions, the class of models we propose could also give rise to some interesting signatures at high-energy colliders. In particular, the exponential suppression of the zero mode couplings implies that quite generically, if the heavier gears can be produced they will decay into dark matter particles with a long lifetime. For our scalar Clockwork model, however, a smoking gun signature is difficult to achieve at the LHC. Since the only interactions the Clockwork sector has with the Standard Model are via the Higgs, there is only a handful of possibilities to access the gears. If the FIMP is lighter than the Higgs ($m_{h} > 2 m_{a_{0}}$), the suppressed coupling $h\to a_{0}a_{0}$ will show up as a small invisible decay width. Since this coupling is of the order of $10^{-10}$ or smaller, measuring this invisible width is beyond the capacity of both the LHC and even a future 100 TeV collider. The second possibility is the production of the heavier gears via an off-shell Higgs. Following the production of the heavier gears, they can decay to the lighter gears through $\mathcal{O} (1)$ couplings. However, for a large enough number of sites, the mass gap between successive states is extremely small, and thus the decay to the standard model via the Higgs will be suppressed by the off-shell nature of the Higgs. Thus the most accessible state is the lightest gear $a_{1}$, with the decay $a_{1}\to h a_{0}$ proceeding via a Clockwork-suppressed coupling. If we consider the scalar FIMP model described in Sec. \ref{scalarcw}, and a diagonal mass of $\rm t~=1~ GeV$, the FIMP obtains a mass of $\rm m_{a_{0}}~\sim 1~GeV$ \footnote{Recall that the Higgs vev only adds a Clockwork-suppressed mass.}, while the lightest gear for q=2 has a mass of $\rm ~\sim 1 ~TeV$ for a Clockwork mass parameter of 1 TeV. While the production of gears via an off-shell Higgs in this mass range is suppressed at the LHC, a 100 TeV collider can potentially access these states.  At a 100 TeV collider, where the cross-section will be large enough, the decay  $a_{1}\to h a_{0}\to b\bar{b} + E_{T}^{miss}$ will produce displaced b-jets in association with missing energy. For the alternative scalar FIMP model described by Eq.\eqref{altscal}, the scalar s is produced via the Higgs and, if it is lighter than half the Higgs mass ($m_{h} > 2 m_{s_{0}}$), would lead to an invisible decay width measurable at both LHC and higher energy colliders. However the higher gears would still be relatively inaccessible at the LHC (potentially accessible at a 100 TeV collider). 

For the fermionic Clockwork described in Sec. \ref{sec:ffimp}, as we argued the gears have to be at a high mass scale, which renders them inaccessible at colliders. In the model described instead by Eq.\eqref{altfer}, since there is no mixing between the Higgs and the Clockwork sectors, the gears can be light. The phenomenology therefore is similar to the scalar Clockwork case.

\section{Summary and discussion}\label{sec:conclusions}

In this paper we studied different ways through which the Clockwork mechanism can be invoked in order to explain the extremely small couplings typically required in freeze-in dark matter production scenarios. We explored two such concrete realisations, one based on a scalar and one on a fermion Clockwork chain and briefly sketched some alternative ideas. In both cases the dark matter candidate is the lightest particle of the Clockwork chain, whose interactions with the Standard Model (as well as with any particle belonging to the same thermal bath as the SM) are feeble due to their suppression by powers of $q^N$, where $q$ is the Clockwork symmetry breaking charge and $N$ the number of sites in the chain. 

We indeed found that for appropriate values of these two parameters the lightest Clockwork states are FIMPs and their freeze-in abundance can meet the one inferred from cosmological observations. We pointed out a simple way to achieve heavier FIMP masses with respect to other similar models proposed in the literature, by introducing a modification of the original Clockwork Lagrangian which does not affect the exponential suppression of the zero mode couplings. We hope this will provide more freedom for model-building ventures along (and beyond) the lines of the simple models we presented. We moreover pointed out several subtleties and computational challenges that can appear in Clockwork FIMP models and proposed ways to tackle them. Lastly, we detailed a method to diagonalise Clockwork-type mass matrices which, to the best of our knowledge, has not been presented as explicitly in the literature before.

Having been motivated primarily by low-energy considerations, one question that we chose not to address in this paper is the dynamical origin of the Clockwork symmetry breaking parameters and the way it could affect our results. For example, if these parameters are viewed as background values of actual dynamical fields, these fields might possess ${\cal{O}}(1)$ couplings with the zero mode and, if they are present in the SM thermal bath after reheating (\textit{i.e.} if they are light enough), they could bring the dark matter candidate in thermal equilibrium with the Standard Model (\textit{cf} also some relevant comments in the ``Clockwork WIMP'' scenario presented in \cite{Hambye:2016qkf}). Such a feature would, clearly, hinder the zero mode from being a viable FIMP dark matter candidate. It would be interesting to study potential UV completions of Clockwork FIMP scenarios, a discussion which we postpone for future work.

\section{Acknowldgements}

We would like to aknowledge enightening discussions with Genevi\`eve Belanger, C\'edric Delaunay, R. Sekhar Chivukula, Ashwani Kushwaha and Alexander Pukhov, whom we would also like to thank for support with {\tt micrOMEGAs5.0}. A.G. was supported by the Labex ILP (reference ANR-10-LABX-63) part of the Idex SUPER, and received financial state aid managed by the Agence Nationale de la Recherche, as part of the programme Investissements d'avenir under the reference ANR-11-IDEX-0004-02. The work of D.S and K.M. is supported by the National Science Foundation, U.S.A under Grant No. 1519045. A. G. and D. S. would like to thank the organisers of the 2017 Les Houches -- Physics at TeV Collders workshop where this project was initiated for their warm hospitality. D.S acknowledges LPTHE-Paris for hospitality during the completion of part of this work. 
K.M. acknowledges LAPTh, Annecy for hospitality where part of this work was carried out.
This work was supported in part by Michigan State University through computational resources provided by the Institute for Cyber-Enabled Research.

\appendix

\section{Diagonalizing tridiagonal and Toeplitz matrices}
\label{sec:diagonalize}

We describe here the procedure to diagonalize general tridiagonal matrices, of which Toeplitz matrices are a special case. We follow \cite{Yueh2005} in the following discussion. A brief proof for a special case along these lines was provided in \cite{Farina:2016tgd}. Consider a $n\times n$ \footnote{In the notation of  the main body 
of the paper, viz. section \ref{sec:clockwork}, n= N+1} tridiagonal matrix 
of the form
\begin{equation}
A_{n} = \begin{bmatrix}
-\alpha + b  & c & 0 & 0 & \dots & 0  & 0\\
a & b & c & 0 & \dots  & 0 &0 \\
0 & a & b & c & \dots & 0 & 0 \\
\dots & \dots  & \dots & \dots & \dots & \dots \\
0 & 0 & 0 & 0 &\dots & a & -\beta + b 
\end{bmatrix}_{n\times n}
\label{tridiag}
\end{equation}
where $ \alpha, a , b, c  \in  \mathbb{C}$.  The eigenvalue equation for this matrix can be written as
\begin{equation}
AX= \lambda X
\end{equation}
where $\lambda$, the eigenvalues, are obtained by solving the characteristic polynomial equation of degree n, $|A-\lambda I |=0 $. 
 We will however proceed to solve the eigenvalue problem 
by a recursion relation of sequences. To this end, we define the sequence $X=\{x\}_{i=0}^{\infty}$, $x_{0}=0$, $x_{n+1}=0$ \footnote{Any quantitiy in curly brackets should be interpreted as a sequence in this section.}, 
such that a recursion relation can be written for any term in the eigenvalue problem, 
\begin{equation}
a x_{i-1} + bx_{i} + cx_{i+1} = \lambda x_{i} + f_{i}, ~i=1,2,\dots,
\end{equation}
with $f_{1} = \alpha x_{1}$, and $f_{n} = \beta x_{n}$, and $f_{i} =0$ for $i \ne 1,n$. Defining a sequence $f=\{f_{i}\}_{i=0}^{\infty}$,
we can rewrite the above recursion relation as
\begin{equation}
c\{x_{i+2} \}_{i=0}^{\infty} + b\{x_{i+1}\}_{i=0}^{\infty} + a\{ x_{i}\}_{i=0}^{\infty} = \lambda\{x_{i+1} \}_{i=0}^{\infty} + \{f_{i+1}\}_{i=0}^{\infty}
\label{recx}
\end{equation}
 
We now further define two sequences $h = \{ 0,1,0,\dots \}$, and $\bar{x} = \{x,0,0,\dots,\}$, and use the Cauchy 
convolution theorem for the product of two sequences, $ a = \{a\}_{i=0}^{\infty}  $,  $ b = \{b\}_{i=0}^{\infty}  $, as
$ab = \sum_{i=0}^{\infty} c_{i} $, where $c_{i} = \sum_{l=0}^{i}a_{l}b_{i-l}$, to rewrite the recursion relation by convolution of Eq.\eqref{recx} with $h^{2}$ as\footnote{To obtain Eq.\eqref{rec3}, note that $h{x_{n+1} = h\{{x_{1},x_{2},\dots}}\} = X- \bar{x_{0}}$. Also note that, $h^{2}\{x_{n+2}\}= X- \bar{x_{0}} - x_{1}h$. Therefore, solving for X, and substituting $x_{0}=f_{0}=0$, we obtain, $(a h^2 + (b-\lambda)h + \bar{c})X = (f + c\bar{x}_{1})h$. Since $c\neq 0$, we obtain Eq. \eqref{rec3}.} 
\begin{equation}
X = \frac{(f + c \bar{x}_{1} ) h}{ ah^{2} + (b - \lambda)h + \bar{c}} \ .
\label{rec3}
\end{equation}
The denominator of this equation has two roots, which can be written as
 \begin{equation}
 \gamma_{\pm} = \frac{ -(b-\lambda) \pm \sqrt{\omega}}{2a}, {\rm ~with~} \omega = (b-\lambda)^{2} - 4 ac ;~ \gamma_{\pm}, \omega \in  \mathbb{C} \ .
 \end{equation}
Let $\gamma = p \pm iq$, such that $\gamma_{+}\gamma_{-} = p^{2} + q^{2}=c/a$, and $\gamma_{+} + \gamma_{-}=2p=\frac{(\lambda-b)}{a}$. We obtain the relations
 \begin{eqnarray}
 \gamma_{\pm} & = & \sqrt{p^{2} + q^{2}}(\cos\theta \pm i\sin\theta)=\frac{1}{\rho}e^{\pm i\theta}  \nonumber\\
 \rho &=&\pm\sqrt{\frac{a}{c}}; \qquad \cos\theta=\frac{p}{\sqrt{p^{2} + q^{2}}}=\frac{\lambda - b}{2\sqrt{ac}}, \qquad \rho, \theta\in \mathbb{C}.
 \label{rel1}
 \end{eqnarray}
We can then decompose the relation in  Eq.\eqref{rec3} in partial fractions as
\begin{equation}
X = \frac{1}{\sqrt{\omega}}\Bigg\{\Bigg( \frac{a}{c}\Bigg)^{j+1}\Bigg(\gamma_{+}^{j+1} - \gamma_{-}^{j+1}\Bigg)\Bigg\}(f + c\bar{x}_{1})h \ .
\end{equation}
Using De-Moivre's theorem we can simplify this expression as
\begin{equation}
X = \frac{2i}{\sqrt{\omega}}\{\rho^{j+1}\sin(j+1)\theta\}(f + c\bar{x}_{1})h \ .
\end{equation}
Since $f_{1}=\alpha x_{1}$,$f_{n}=\beta x_{n}$, and $f_{j}=0$ for $j\ne 1,n$, we can obtain an expression for $x_{j}$
\begin{equation}
x_{j} =\frac{2i}{\sqrt{\omega}}(c x_{1}\rho^{j}\sin(j\theta) + \alpha x_{1} \rho^{j-1}\sin (j-1)\theta + H(j-n-1)\beta x_{n}\rho^{j-n}\sin (j-n)\theta  )
\end{equation}
where H(x), the Heaviside step function is defined the usual way: $H(x)=1$ if $\rm x \geq 0$, and $H(x)=0$ if $x < 0$. The crucial recurence relation is
\begin{equation}
\frac{\sqrt{\omega}}{2i}x_{j+1}= cx_{1}\rho^{j+1}\sin(j+1)\theta + (\alpha + \beta)x_{1}\rho^{j}\sin j\theta + \frac{1}{c}\alpha\beta x_{1}\rho^{j-1}\sin (j-1)\theta
\end{equation}
Since $\rho,x_{1}\ne 0$, and $x_{j+1}=0$, we obtain the definitive condition
\begin{equation}
\boxed{ac\sin (n+1)\theta + (\alpha + \beta)\rho c \sin n\theta + \alpha\beta \sin (n-1)\theta =0} 
\label{rel2}
\end{equation}
From Eq.\eqref{rel1}, the eigenvalues are given by
\begin{equation}
\boxed{\lambda = b + \frac{2a}{\rho}\cos\theta}
\label{eigval}
\end{equation}
We are, thus, left with solving Eq.\eqref{rel2} for $\theta$ to obtain the eigenvalues $\lambda$ given by Eq.\eqref{eigval}.  
The eigenvectors, on the other hand, are given by
\begin{equation}
\boxed{x_{j} = \frac{x_{1}\rho^{(j-1)}}{\sin\theta}[\sin j\theta + \frac{\alpha}{\rho c} \sin(j-1)\theta]}
\label{eigvec}
\end{equation}
For the Clockwork matrix $b=1+q^{2}$, $\alpha=q^{2}$, $\beta=1$, $a=c=-q$, $\rho=\pm 1$.
Thus the condition in Eq.\eqref{rel2} reduces to,  
\begin{equation}
q^{2}(2\sin n\theta\cos\theta) - q(1+q^{2})\rho\sin n\theta=0
\label{rel3}
\end{equation}  
yielding the conditions
\begin{equation}
\rm \sin n\theta =0, ~or~\cos\theta =\frac{1}{2q}(1+q^{2})\rho \ .
\end{equation}
For $\sin n\theta=0 \Rightarrow \theta=\frac{k\pi}{n}$, yielding the $n-1$ eigenvalues, for $\rho=+1$,
\begin{equation}
\boxed{\lambda_{k}= (1+q^{2}) -2q\cos\frac{k\pi}{n},~k=1,\dots,n-1}
\end{equation}
The eigenvectors are obtained from Eq.\eqref{eigvec}, by choosing $x_{1}=\sin\theta$
\begin{equation}
\boxed{(x_{j})_{k}= Z \Bigg[ \sin\frac{jk\pi}{n} - q\sin\frac{(j-1)k\pi}{n} \Bigg],~j=1,\dots,n;~k=1,\dots,n-1}
\end{equation} 
where $Z$ is a normalization factor found to be $Z=\sqrt{\frac{2}{n\lambda_{k}}}$.
\\
The remaining eigenvalue is obtained from the condition $\cos\theta = \frac{1}{2q}(1+q^{2})\rho $, yielding
\begin{equation}
\boxed{\lambda=0} \ .
\end{equation}
It corresponds to the zero mode of the Clockwork matrix. To obtain the corresponding eigenvector we first note that $\sin\theta=\frac{i(1-q^{2})}{2q}$ and $e^{i\theta}=q$. Hence, from Eq.\eqref{eigvec}, and eventually choosing $x_{1}=1$, we get
\begin{eqnarray}
(x_{j})_{0} &=& Z\frac{x_{1}}{\sin\theta}\Bigg[\sin j\theta - q\sin(j-1)\theta\Bigg] \nonumber \\
 & = & Z \frac{x_{1}}{\sin\theta}\Bigg[\frac{e^{ij\theta} - e^{-ij\theta}}{2i} - \frac{e^{i(j-1)\theta} - e^{-i(j-1)\theta}}{2i}\Bigg] \nonumber\\
 &\Rightarrow & \boxed{(x_{j})_{0} = \frac{Z}{q^{j-1}}}
\end{eqnarray}
with the normalization factor being $Z=\sqrt{\frac{q^{2}-1}{q^{2}- q^{-2(n-1)}}}$. 

We can now attempt to deform the Clockwork matrix. To this end, we first note that this matrix is rather special, as $\alpha,\beta,a,c$ have particular values that allow for an analytic solution of Eq.\eqref{rel2}. If $\alpha=\beta=0 $, the matrix      Eq. \eqref{tridiag} reduces to a Toeplitz matrix, which can be easily diagonalised by solving Eq.\eqref{rel2}. Next we note that adding a constant diagonal matrix to Eq.\eqref{tridiag}, the eigenvalues just get shifted by the constant diagonal, given by b, while the eigenvectors do not change at all. For arbitrary values of $\alpha$, and $\beta$, the transcendental Equation \eqref{rel2} has to be solved numerically. We have however checked numerically that the $\beta$-dependence  of the eigenvalues and the eigenvectors is minimal. The zero mode is, thus, only corrected by a very small amount for $\beta > 1$, with the eigenvector remaining the same. Essentially, any non-zero value for $\beta$ is ``Clockworked'' and the zero mode remains nearly massless. The other eigenvalues are corrected by order one numbers.  Changing $\alpha > 1$, however, has the opposite effect. Numerical diagonalization reveals that the zero mode gets corrected by an order one number, and that the ``Clockwork mechanism'' does not work anymore.   

\section{Diagonalizing the fermionic Clockwork matrix} 
\label{sec:fcwdiag} 
To diagonalise the mass matrix in Eq.\eqref{fcwm}, we rewrite it as
\begin{equation}
\mathcal{M}  = \begin{bmatrix}
A_{l\times l} & B_{l\times r} \\
B^{T}_{r\times l} & D_{r\times r}
\end{bmatrix}_{(2n+1)\times(2n+1)}
\end{equation}
where, for $l = n, r = n+1$, the matrices $A$, $D$ and $B$ read
\begin{eqnarray}
A & = & (\mathds{I}_{l\times l})m\tilde{q},~D=(\mathds{I}_{r\times r})m\tilde{q}  \nonumber\\
& & \nonumber \\
B & = & m \begin{bmatrix}
1 & -q  & 0  & 0 & \dots & 0 \\
0 &  1  & -q & 0 & \dots & 0 \\
0 &  0  &  1 & -q & \dots & 0 \\
\vdots & \vdots & \vdots & \vdots & \vdots & \vdots \\
0 & 0 & 0 & 0 & 1 & -q \\ 
\end{bmatrix}_{n\times n+1} \ .
\end{eqnarray}
We first note that the matrix $(B^{T}B)_{r\times r}$ is given by 
 \begin{equation}
 (B^{T}B)_{r\times r} = \begin{bmatrix}
1 & -q & 0 & 0 &\dots & 0  & 0\\
-q & 1+ q^{2} & -q & 0 & \dots & 0 & 0 \\
0 & -q & 1 + q^{2} & -q & \dots & 0 & 0 \\
\vdots & \vdots & \vdots & \vdots & \vdots & \vdots & \vdots \\ 
0 & 0 & 0  & 0 &\dots & 1+ q^{2} & -q \\
0 & 0 & 0 & 0 & \dots & -q & q^{2}
\end{bmatrix}_{(n+1)\times(n+1)}
\label{scwm}
 \end{equation}
which is exactly equal to the scalar Clockwork matrix, with eigenvalues and eigenvectors given by Eqs.\eqref{scweig} and \eqref{scweigvec}. We also note that the matrix $(BB^{T})$ is given by
 \begin{equation}
 (BB^{T})_{l\times l} = \begin{bmatrix}
1 + q^{2} & -q & 0 & 0 &\dots & 0 & 0 \\
-q & 1+ q^{2} & -q & 0 & \dots & 0 & 0 \\
0 & -q & 1 + q^{2} & -q & \dots & 0 & 0 \\
\vdots & \vdots & \vdots & \vdots & \vdots & \vdots & \vdots \\ 
0 & 0 & 0  & 0 &\dots & 1+ q^{2} & -q \\
0 & 0 & 0 & 0 & \dots & -q & 1+ q^{2}
\end{bmatrix}_{n\times n}
 \end{equation}
which is a banded ($n\times n$) Toeplitz matrix with eigenvalues and eigenvectors given by Eq. \eqref{eigval},
 \begin{eqnarray}
 \lambda_{k} &=  & 1+ q^{2} - 2\cos \frac{k\pi}{(n+1)},~(k=1,\dots ,n),   \nonumber\\ 
 \chi_{k} &=& \sqrt{\frac{2}{n+1}}\sin\frac{ik\pi}{n+1},~~(i=1,\dots ,n)
 \end{eqnarray}
To solve the eigenvalue equation, $\mathcal{M} U = \Lambda U$, we write
\begin{equation}
U= \begin{bmatrix}
X \\
Y
\end{bmatrix}  
\end{equation}  
where $X$ and $Y$ are column  vectors of size l and  r respectively. Then, the eigenvalue equation is
\begin{equation}
\begin{bmatrix}
A_{l\times l} & B_{l\times r} \\
B^{T}_{r\times l} & D_{r\times r}
\end{bmatrix}
\begin{bmatrix}
X_{l} \\
Y_{r}
\end{bmatrix} =
\Lambda
\begin{bmatrix}
X_{l} \\
Y_{r}
\end{bmatrix} 
\end{equation}
 which leads to two equations, 
 \begin{eqnarray}
 (A_{l \times l} - \Lambda\mathds{I}_{l \times l})X_{l} + B_{l\times r}Y_{r}  & = & 0 \nonumber \\
 B^{T}_{r\times l} X_{l} +  (D_{r \times r} - \Lambda\mathds{I}_{r\times r})Y_{r} & = & 0 
 \end{eqnarray}
 
 Since A and D are diagonal, we have the trivial relations, 
 \begin{eqnarray}
 (A - \Lambda)_{l\times l}  &= & (m\tilde{q} - \Lambda)\mathds{I}_{l \times l} ; ~ (D - \Lambda)_{r\times r}  =  (m\tilde{q} - \Lambda)\mathds{I}_{r \times r} \nonumber \\
 (A - \Lambda)_{l\times l}^{-1}  &= & \frac{1}{(m\tilde{q} - \Lambda)}\mathds{I}_{l \times l} ; ~ (D - \Lambda)_{r\times r}^{-1}  =  \frac{1}{(m\tilde{q} - \Lambda)}\mathds{I}_{r \times r} 
 \end{eqnarray}
 This yields the first eigenvalue, $\Lambda = m\tilde{q}$.
Additionally, using the above set of equations, we obtain the following equations for the eigenvalue  problem
\begin{equation*}
[(B^{T}B)_{r\times r}   +  (m\tilde{q} - \Lambda)^{2}\mathds{I}_{r\times r}]Y_{r}   =  0  
\end{equation*}
\begin{equation}
[(BB^{T})_{l\times l }   +  (m\tilde{q} - \Lambda)^{2}\mathds{I}_{l\times l}]X_{l}   =  0  \ .
\label{fcweig-vec}
\end{equation}
Let S and T be the matrices that diagonalize $B^{T} B$ and $BB^{T}$ respectively. Then we obtain
\begin{equation}
T^{-1}[(BB^{T})_{l\times l} + (m\tilde{q} -\Lambda)^{2}\mathds{I}_{l\times l}]T =0 \ .
\end{equation}
The eigenvalues of $BB^{T}$ have already been calculated, and thus the eigenvalue for the above equation is 
\begin{equation}
\Lambda_{k} = m\tilde{q} \pm m\sqrt{\lambda_{k}} \ .
\end{equation}  
Similarly, solving the second condition for $B^{T}B$ yields the second set of eigenvalues, and thus we obtain the set, 
\begin{equation}  
\boxed{\Lambda_{k} = m\tilde{q}, m\tilde{q} \pm m\sqrt{\lambda_{k}}}
\end{equation}
In order to find the eigenvectors, we note that the eigenvector blocks $X,Y$ should simultaneously satisfy Eq.\eqref{fcweig-vec}. For the eigenvector $\Lambda_{k}=m\tilde{q}$, while $Y_{q} = \frac{\mathcal{N}_{0}}{q^{j}},~(j=1,\dots,N)$, the second of the equations in Eq.\eqref{fcweig-vec} can only be satisfied for $X=\vec{O}$, the null vector. This yields part of the eigenvectors.  Noting that the fermionic mass matrix is diagonalised by a bi-unitary transformation in the $\psi_{L}-\psi_{R}$ basis , we make an ansatz for the eigenvectors and write them as
\begin{equation}
\mathcal{U} =
 \begin{bmatrix}
\vec{O} &  \alpha U_{L} &  \beta U_{L} \\
\vec{u}_{R} & \gamma U_{R} &   \delta U_{R} \\
\end{bmatrix}
\end{equation}
where $u_{R},U_{L},U_{R}$ are given by Eq.\eqref{fcweigvec}. The constants and the signs of the constants $\alpha,\beta,\gamma,\delta$ are to be fixed by normalization and orthogonalization of the eigenvectors $\mathcal{U}^{T}\mathcal{U} =\mathds{1}$. This yields, $\alpha=\gamma=\delta=\frac{1}{\sqrt{2}};~\beta=-\frac{1}{\sqrt{2}}$ and thus the eigenvectors are given by Eq.\eqref{fcweigvec}.

\bibliographystyle{utphys}
\bibliography{ref.bib}
\end{document}